\documentclass[12pt]{article}
\usepackage{graphicx}
\usepackage{pazh_win1}

\usepackage{lscape}
\usepackage{amssymb}
\usepackage{amsmath}

\usepackage[T2A]{fontenc}
\usepackage[english]{babel}

\AtBeginDocument{%
  
}

\AtBeginDocument{%
  
}

\AtBeginDocument{%
  
}

\graphicspath{{figs/}}

\usepackage{natbib}

\usepackage{color}

\tightenlines

\begin{document}

\title{Prospects of observation of gravitationally lensed sources by space submillimeter telescopes.}

\author{\hspace{-1.3cm}\copyright\, 2019 \ \
 T.I. Larchenkova\affilmark{1}$^{\,*}$,
 A.A. Ermash\affilmark{1}$^{\,}$,
 A.G. Doroshkevich\affilmark{1}$^{\,}$}
\affil{$^1$ {\it Lebedev Physical Institute of Russian Academy of Sciences, Moscow, Russia.}\\}

\date{\today}

%\vspace{5mm}

\sloppypar \vspace{2mm}

\begin{abstract}
\noindent
In the current paper we consider the prospects of observation of gravitationally lensed sources in far-infrared and submillimeter wavelength ranges by
the future space telescopes equipped with actively cooled main mirrors.
We consider the possibility of solving some of the important cosmological and astrophysical scientific tasks by observation of gravitationally lensed sources.
The number counts of lensed sources for wavelengths from 70$\mu m$ up to 2000$\mu m$ were calculated.
We discuss the distribution of lensed sources at different redshifts and magnification coefficients and also mass distribution of lenses.
Model sky maps that illustrate the contribution of lensed sources were created.
\end{abstract}

{\it Keywords:\/} far IR, galaxy evolution, gravitational lensing.

{$^{*}$ E-mail: ltanya@asc.rssi.ru}

\section*{INTRODUCTION}
\renewcommand{\baselinestretch}{1.0}

Modern astrophysical instruments with their growing angular resolution and sensitivity make it possible to observe more and more sources that undervent 
strong gravitational lensing.
The gravitational lensing occurs when the light from a distant background object is deflected by a massive object - gravitational lens on the line of sight.
Any object can act as a gravitational lens, e.g. galaxy or galaxy group.
In the case of strong gravitational lensing multiple images of the source can be observed.
Angular distance between such images of a lensed objects increases with the lens mass if the impact parameter remains constant.
That is why it is easier to detect lensed systems where lens is a high-mass galaxy or a galaxy cluster.
The first discovered strongly lensed object was the quasar B$0957 + 561$ with two images that was lensed by a massive galaxy \citep{1979Natur.279..381W}.
The information about the time delay between the images; their fluxes and coordinates makes it possible to recover the mass distribution in the lens and also
to estimate the main cosmological parameters.

If an extended source is lensed, e.g. a galaxy, an Einstein-Chwolson ring or a part of it might be observed.
If a massive galaxy cluster acts as a lens, galaxies beyond it are observed as bright extended arcs.
Such arcs were discovered in the galaxy cluster A370 \citep{1987A&A...172L..14S,1986BAAS...18R1014L} and interpreted in \cite{1987Natur.325..572P} as images of lensed background galaxies.

The amount of known gravitationally lensed systems is limited.
For example, in the CASTLES\footnote{www.cfa.harvard.edu/castles} catalog  there are about a hundred of objects with multiple images.
Such objects provide crucial information for a number of cosmological and astrophysical tasks.

One of such tasks is the independent determination of the Hubble constant $H_0$ (see, e.g., \citealt{1964MNRAS.128..307R,2017MNRAS.468.2590S}).
It is one of the main cosmological parameters that describe the velocity of expansion, age, size and critical density of the Universe.
Other cosmological parameters beside the Hubble constant can also be constrained (see, e.g. \citealt{1992ARA&A..30..311B}).
Strong gravitationally lensed sources allow us to observe the very high redshift objects due to significant flux magnification.
The information of the lenses themselves (galaxies and galaxy clusters) can also be obtained.
One can derive dark matter distribution (see, e.g., \citealt{1991ApJ...373..354K}), initial mass function, SMBH evolution and the influence of the AGN feedback \citep{2006ApJ...649..616P}.
Some papers are dedicated to the low-mass subhalo detection, see the pioneering work by \cite{1998MNRAS.295..587M}.

Instruments that have sufficient sensitivity and angular resolution make it possible to explore the aforementioned problems in various wavelength ranges,
including far-infrared (FIR), submillimeter and millimeter wavelengths.
There are several future space observatories with actively cooled main mirrors that will perform observations in this wavelength range, such as
SPICA\footnote{spica-mission.org} and OST\footnote{asd.gsfc.nasa.gov/firs}, but 
Millimetron is the only one that is expected to be launched in the next decade~\citep{2012SPIE.8442E..4CS,2014PhyU...57.1199K,2017ARep...61..310K}.
Millimetron can perform observations dedicated to all of the aforementioned scientific tasks, including Lyman-alpha emitters, first galaxies and high-redshift 
dusty starforming galaxies (DSFGs) (see review by \citealt{2014PhR...541...45C}).

Aside from using gravitational lensing for tasks of astrophysics and cosmology it plays crucial role in modeling of the extragalactic background.
The gravitational lensing significantly affects number counts in submillimeter and millimeter wavelengths due to the steepness of the curve
(see, e.g., \citealt{2011A_and_A...529A...4B,2017AstL...43..644P}).
Observations in these wavelength ranges showed that the number of sources with large fluxes varies for different sky areas.
This effect is caused by gravitational lensing of sources on galaxies and galaxy clusters (\citealt{2011MNRAS.415.3831A}), see also discussion in \cite{2013MNRAS.428.2529H}.

In this paper we discuss the possible contribution of the Millimetron mission to these scientific tasks.
The structure of the paper is as follows.
In section~\ref{sec:model_parameters} we give the information about the parameters of the Millimetron mission and the description of the model
used to estimate the parameters of lensed objects. 
In section~\ref{sec:model_maps_and_counts} we estimate the number of lensed sources for different wavelengths, their contribution to the number counts,
redshift distribution of lensed sources, magnification coefficient distribution and mass distribution.
Model maps that illustrate the contribution of lensed sources are also given.
In section~\ref{sec:science_task} we consider important astrophysical tasks and the expected parameters of gravitationally lensed systems.
In section~\ref{sec:discussion} we describe the obtained results  and give a short discussion.

\section{The main parameters of the model and the Millimetron space observatory}
\label{sec:model_parameters}

The model that was used to perform all calculations in this paper was described in detail in~\cite{2018arXiv181208575E}.
We have used the numerical model of the dark matter evolution from~\cite{2015A&A...575A..32C}. 
The simulation contained $1024^3$ particles with mass $8.536\times10^{7}M_{\odot}$ in volume $\sim150$~Mpc$^{3}$.
The parameters of the Cosmology were as following.
Matter density $\Omega_m = 0.24$, dark energy density $\Omega_{\Lambda} = 0.76$, baryon fraction $f_b = 0.16$,
dimensionless Hubble constant $h = 0.73$.
Minimum halo mass was $1.7\times10^{9}M_{\odot}$.
We created the cone from the simulation following the approach by~\cite{2005MNRAS.360..159B}.  
Each cube was affected by a random transformation: turn by $-\pi/2$, $0$ or $+\pi/2$, random shift and  mirroring independently on each axis.
Such an approach helps to overcome the effect of perspective.
Next we create a SED library using the GRASIL code~\citep{1998ApJ...509..103S}.

The main difference between the model used in this paper \citep{2018arXiv181208575E} and
our previous model published in \cite{2017AstL...43..644P} is that we used this model SED (Spectral Energy Distribution) library.
Each galaxy was assigned a closest in parameter space model SED (age of galaxy, stellar mass, mass of gas, star formation rate, size and metallicity).
For each disk and bulge 10 SEDs for different inclination angles was created ($0^o$, $10^o$ etc).
Inclination angle of each galaxy was chosen randomly. We utilized the AGN SED from~\cite{2017ApJ...841...76L}.

Each object of the simulation was assigned a SED according to its physical parameters.
Then we calculated the number counts, created model maps for various wavelengths etc. 

Model maps and number counts are significantly affected by the effect of lensing of distant objects.
It becomes more apparent on larger wavelengths.
Earlier \cite{2010Sci...330..800N, 2011A_and_A...529A...4B, 2017AstL...43..644P} showed that the expected fraction of lensed sources at 500$\mu m$ with flux greater that 100mJy is about 15\% and 
reaches 40\% at 1000$\mu m$.
Two simple models of a gravitational lens were considered in this paper: point lens and singular isothermic sphere (see, e.g. \citealt{1992grle.book.....S}).
Only events with magnification greater than or equal to 2 will be considered in this paper.

The main parameters of the Millimetron observatory are given in~\cite{2012SPIE.8442E..4CS,2014PhyU...57.1199K,2017ARep...61..310K} and on the official web-site of the project\footnote{millimetron.ru}.  
It will have an actively cooled to 10K main mirror.
Photometric observations will be performed by the Long wave Array Camera Spectrometer (LACS) 
and Short wave Array Camera Spectrometer (SACS).
List of the wavelength bands and the corresponding FWHMs are given in table~\ref{tab:millimetron_detectors}. 
At $\lambda>300\mu m$ angular resolution is limited by diffraction, and at the shortest band it will be about 1-2 angular seconds.
The expected sensitivity of the wide band photometry is about 0.01mJy.
The field of view of the telescope is $6'\times 6'$.

\begin{table}
\centering
\caption{Parameters of the LACS and SACS detectors of the Millimetron observatory}
\label{tab:millimetron_detectors}
\vspace{2mm}
\begin{tabular}{c|c|c}
\hline
Wavelength ($\mu m$) & FWHM ('') \\
\hline
\multicolumn{3}{c}{Long wave Array Camera Spectrometer (LACS)}\\
\hline
Band 1 & 3000--1500 &  42  \\
Band 2 & 1500--850  &  22  \\
Band 3 & 850--450   &  12  \\
Band 4 & 450--300   &  7.5 \\
\hline
\multicolumn{3}{c}{Short wave Array Camera Spectrometer (SACS)}\\
\hline
Band 1 & 50 --90  & 1--2  \\
Band 2 & 90 --160 & 2--4  \\
Band 3 & 160--300 & 4--6  \\
Band 4 & 300--450 & 6--10  \\
\hline
\end{tabular}
\end{table}

\section{Model maps and lensed source counts}

\label{sec:model_maps_and_counts}
In order to approach scientific tasks connected with gravitationally lensed systems photometric and spectoscopic observations are requied.
The goal of photometric observations is the detection of galaxies and AGN on observational maps and study of their physical properties and evolution.
The number counts is a widely used statistical description of the observed sources.
They can be represented in integral $N(>S)$[deg$^{-2}$] or differential $dN/dS$[mJy$^{-1}$deg$^{-2}$] form.
The number counts can be used to derive the background intensity and to make some estimates of the confusion noise.
This effect takes place when there are more than one source per beam.
Far infrared and submillimeter observations are significantly affected because of the limited angular resolution.
The confusion limit for the Millimetron mission is shown in table~\ref{tab:confusion limit} for a list of wavelengths.
Our estimations show that, e.g. at 110$\mu m$ confusion is  $\sim 10^{-2}$ mJy, that is significantly better that the
confusion limit of the Hershel observatory.

\begin{table}
\caption{Confusion limit estimations for the Millimetron mission.}
\label{tab:confusion limit}
\vspace{2mm}
\begin{tabular}{c|c|c|c}
\hline
Wavelength  & Source density        & Photometry             & Probability of                 \\
($\mu m$)   & criterion (mJy)       & criterion $q=5$ (mJy)  & deflection P(D) $5\sigma$(mJy) \\
\hline
70      &    1.8$\times10^{-3}$     &   6$\times10^{-4}$     &    1.9$\times10^{-3}$    \\
110     &    3.50$\times10^{-2}$    &   3.6$\times 10^{-3}$  &    2.57$\times 10^{-2}$  \\
250     &    2.19                   &   1.44$\times 10^{-1}$ &    2.18                  \\
350     &    2.91                   &   3.90                 &    3.99                  \\
650     &    1.62                   &   3.44                 &    3.37                  \\
850     &    1.01                   &   2.41                 &    2.34                  \\
1100    &    6.06$\times10^{-1}$    &   1.59                 &    1.51                  \\
2000    &    1.66$\times10^{-1}$    &   4.63$\times10^{-1}$  &    4.30$\times 10^{-1}$  \\
\hline
\end{tabular}
\end{table}

\begin{figure*}
\centering
\includegraphics[width=0.7\textwidth]{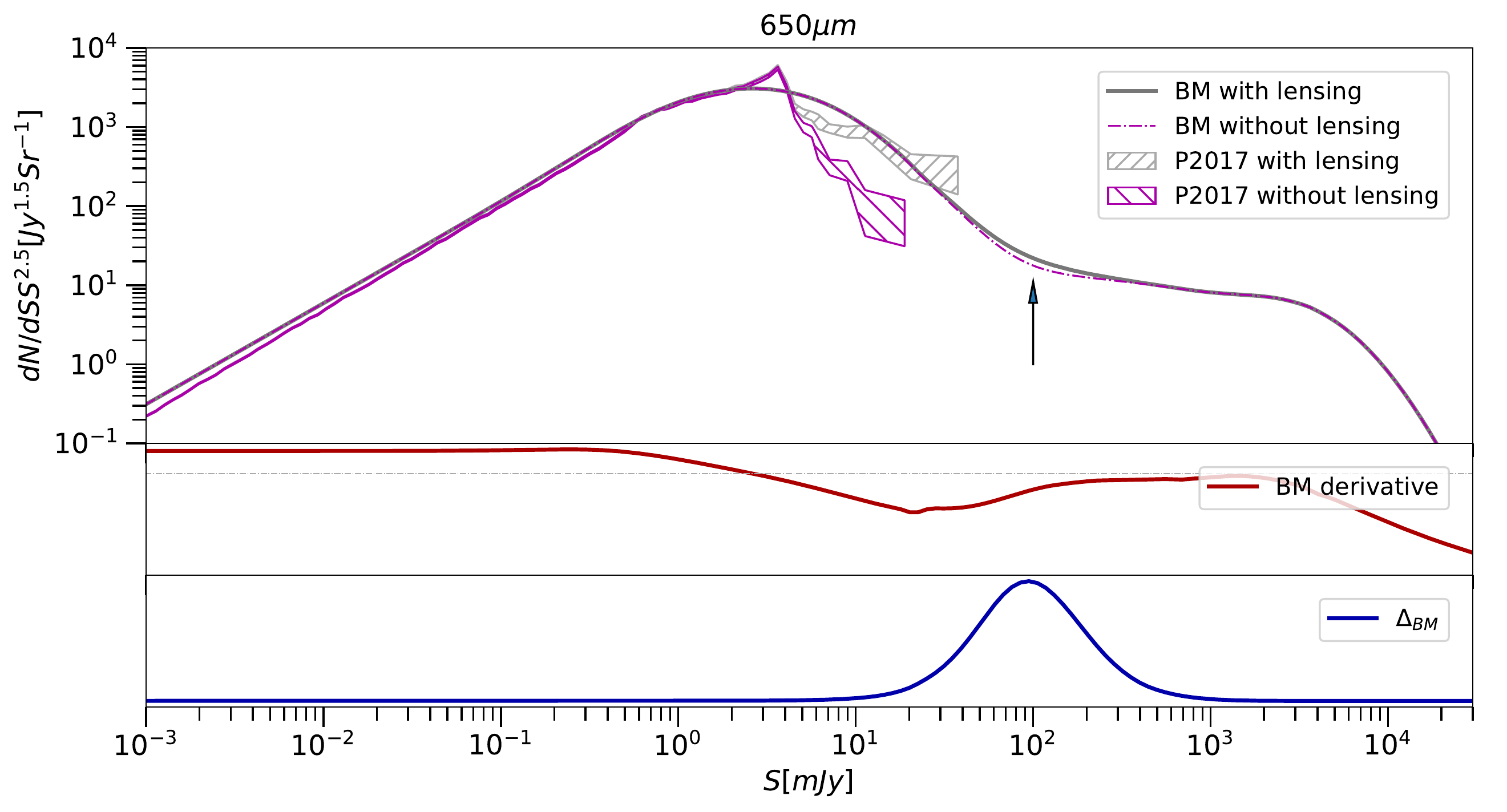}
\caption{Upper panel - differential number counts at $650\mu m$ of the BM and P2017 models with and without lensing,
middle panel~-- steepness of the number counts curve,
lower panel~-- change of the number counts due to the lensing.}
\label{fig:diff_counts_0650}
\end{figure*}

As was shown earlier (see, e.g., \citealt{2011A_and_A...529A...4B,2017AstL...43..644P,2018arXiv181208575E}) gravitational lensing significantly changes the shape of the number counts curve.
This effect is most prominent at millimeter and submillimeter wavelengths at fluxes larger than $\geq 100$ mJy.
This effect can be seen on fig.~\ref{fig:diff_counts_0650}.
The upper panel shows differential number counts of the BM \citep{2011A_and_A...529A...4B} and P2017 \citep{2017AstL...43..644P} models,
central panel shows steepness of the curve for the BM model,
lower panel~-- change in number counts due to the lensing effect.

\subsection{Gravitationally lensed sources on model maps}

\begin{figure*}
\includegraphics[width=0.5\textwidth]{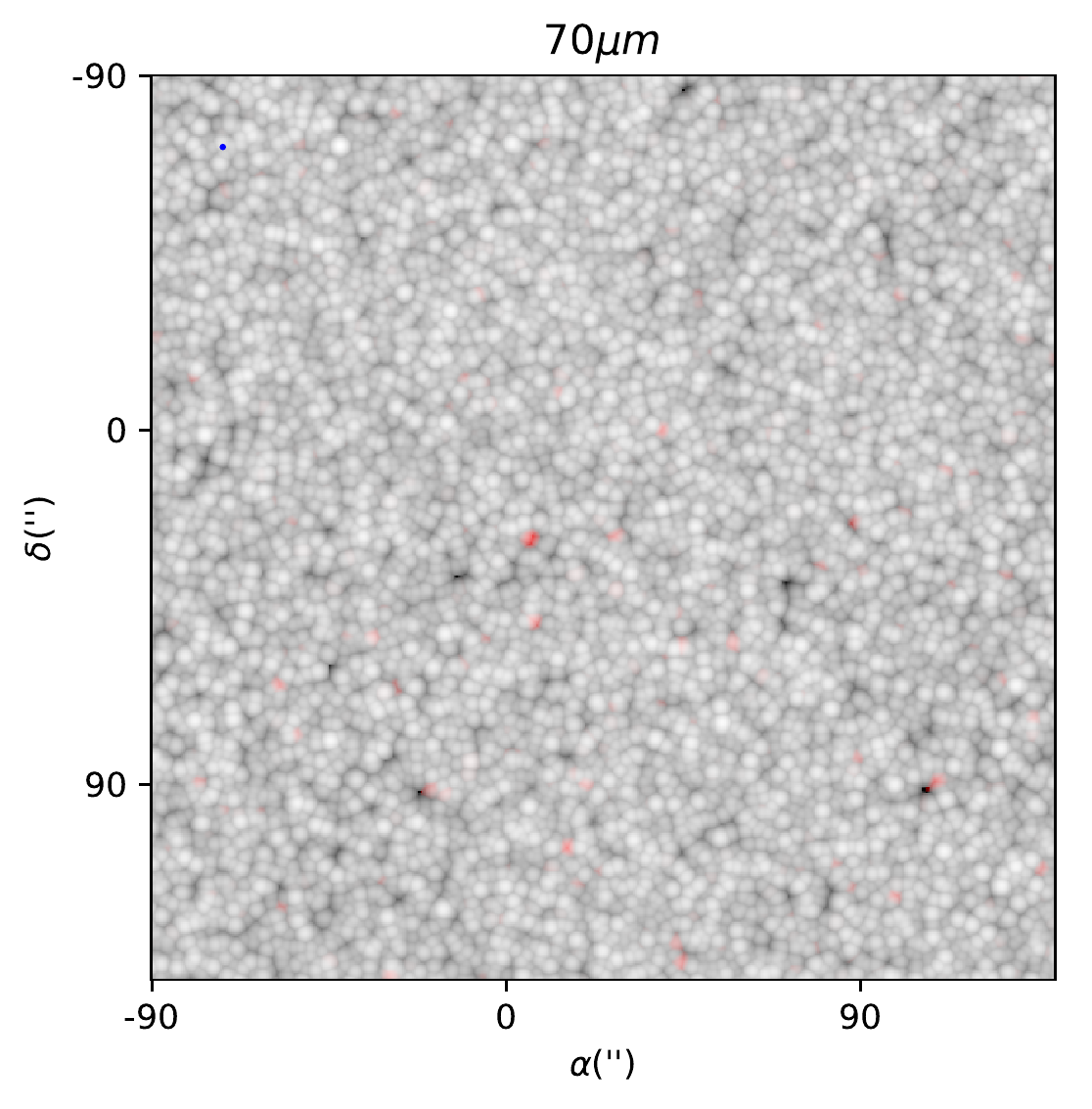}
\includegraphics[width=0.5\textwidth]{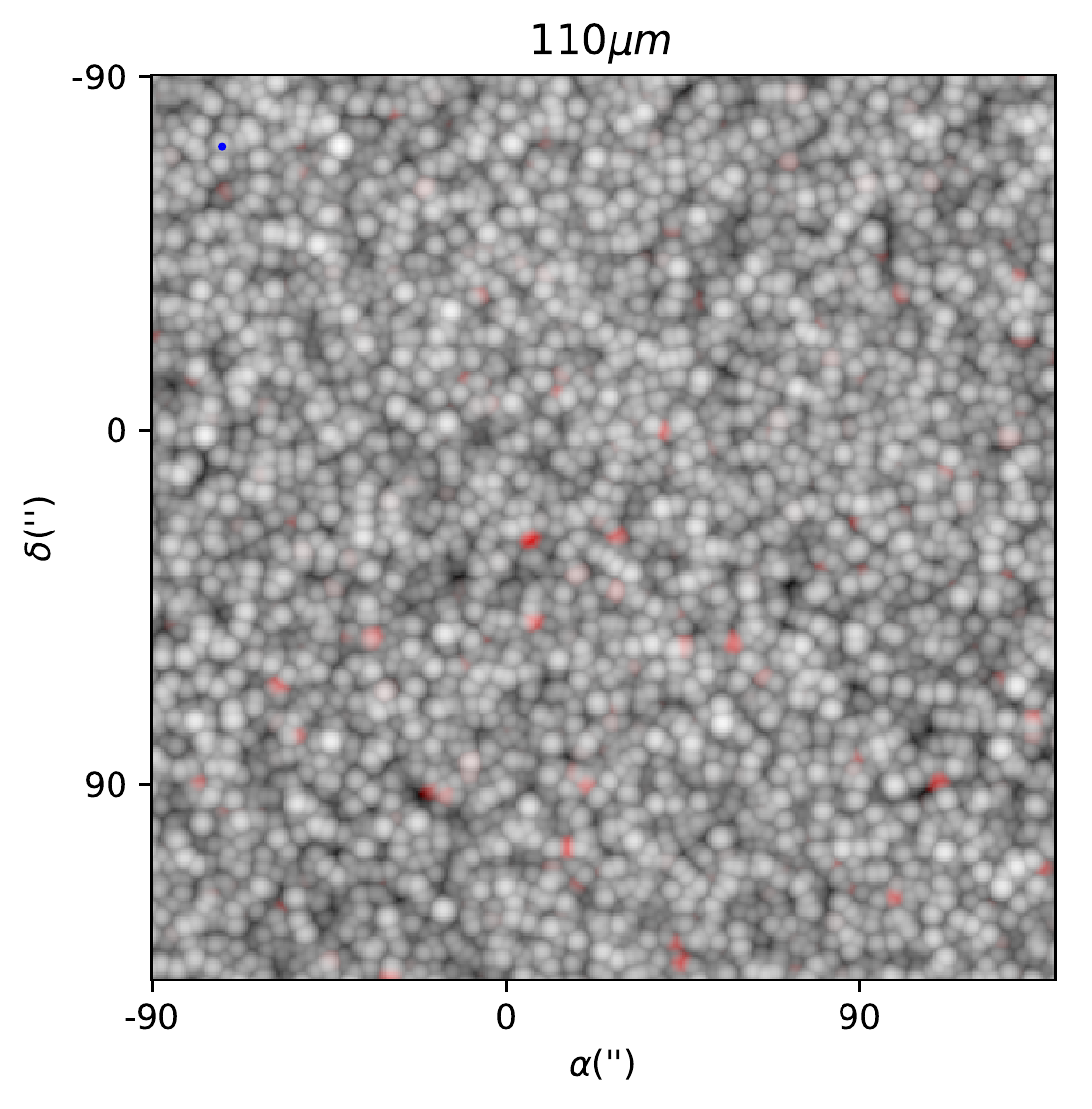}\\
\includegraphics[width=0.5\textwidth]{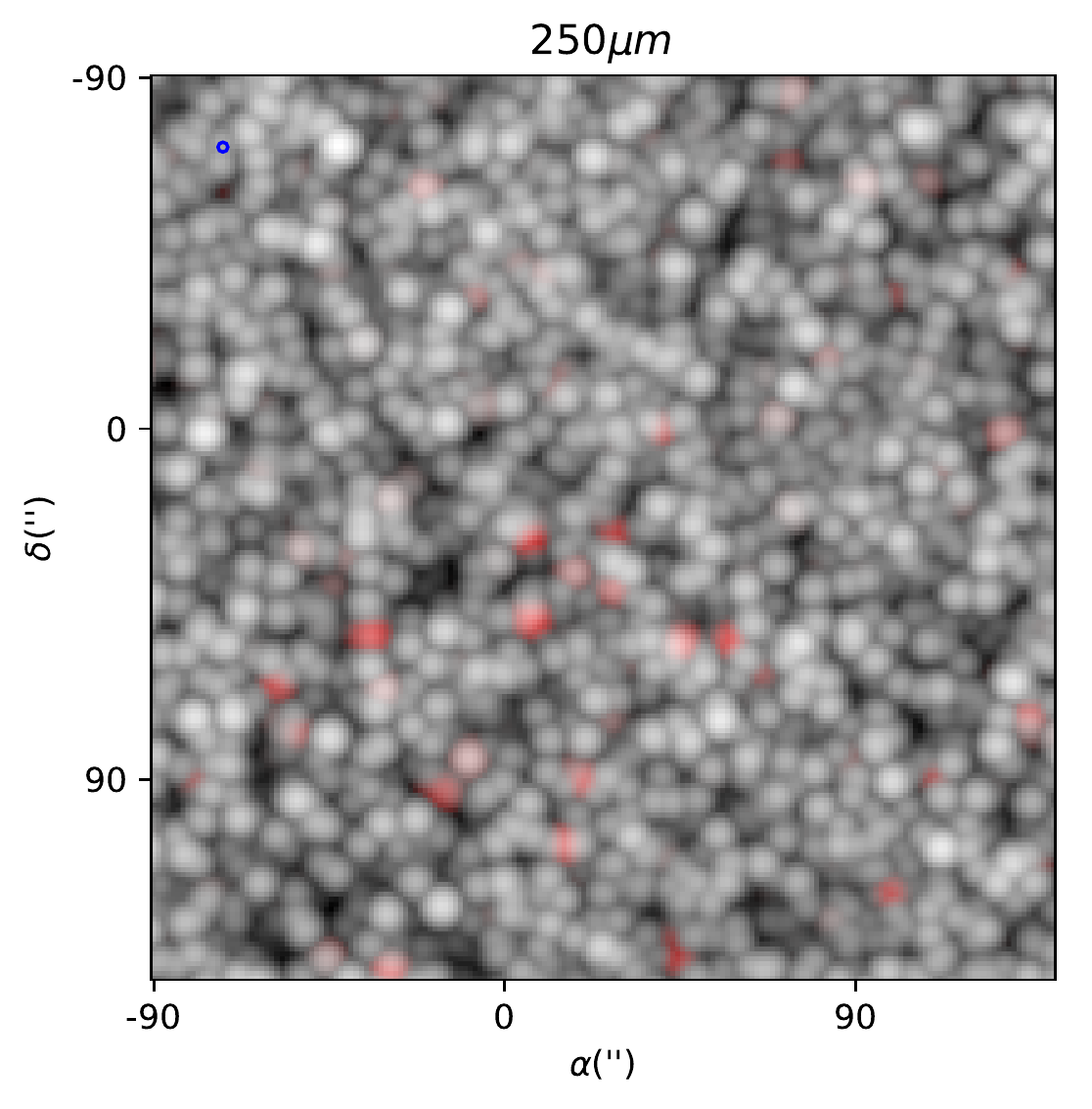}
\includegraphics[width=0.5\textwidth]{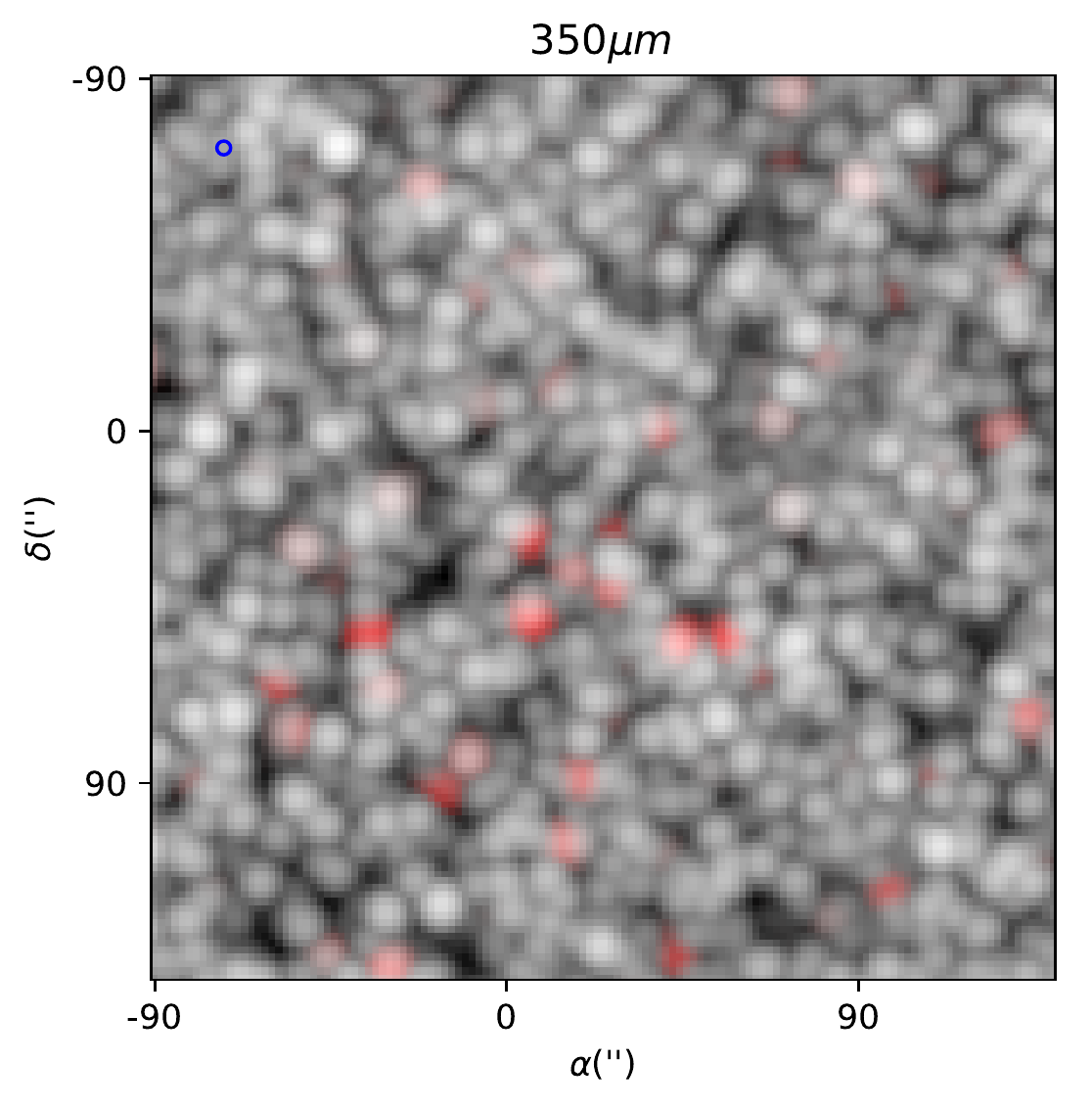}
\caption{Model maps of the $3'\times 3'$ area for 70, 110, 250 and 350 $\mu m$.
Angular resolution corresponds to the diffraction limit of a $10m$ mirror, pixel size is FWHM/3.
Blue circle at the top left of every map shows the beam size.
Lensed sources with magnification $\geq$ 2  are depicted as red. Detailed description see in text.}
\label{fig:maps1}
\end{figure*}

\begin{figure*}
\includegraphics[width=0.5\textwidth]{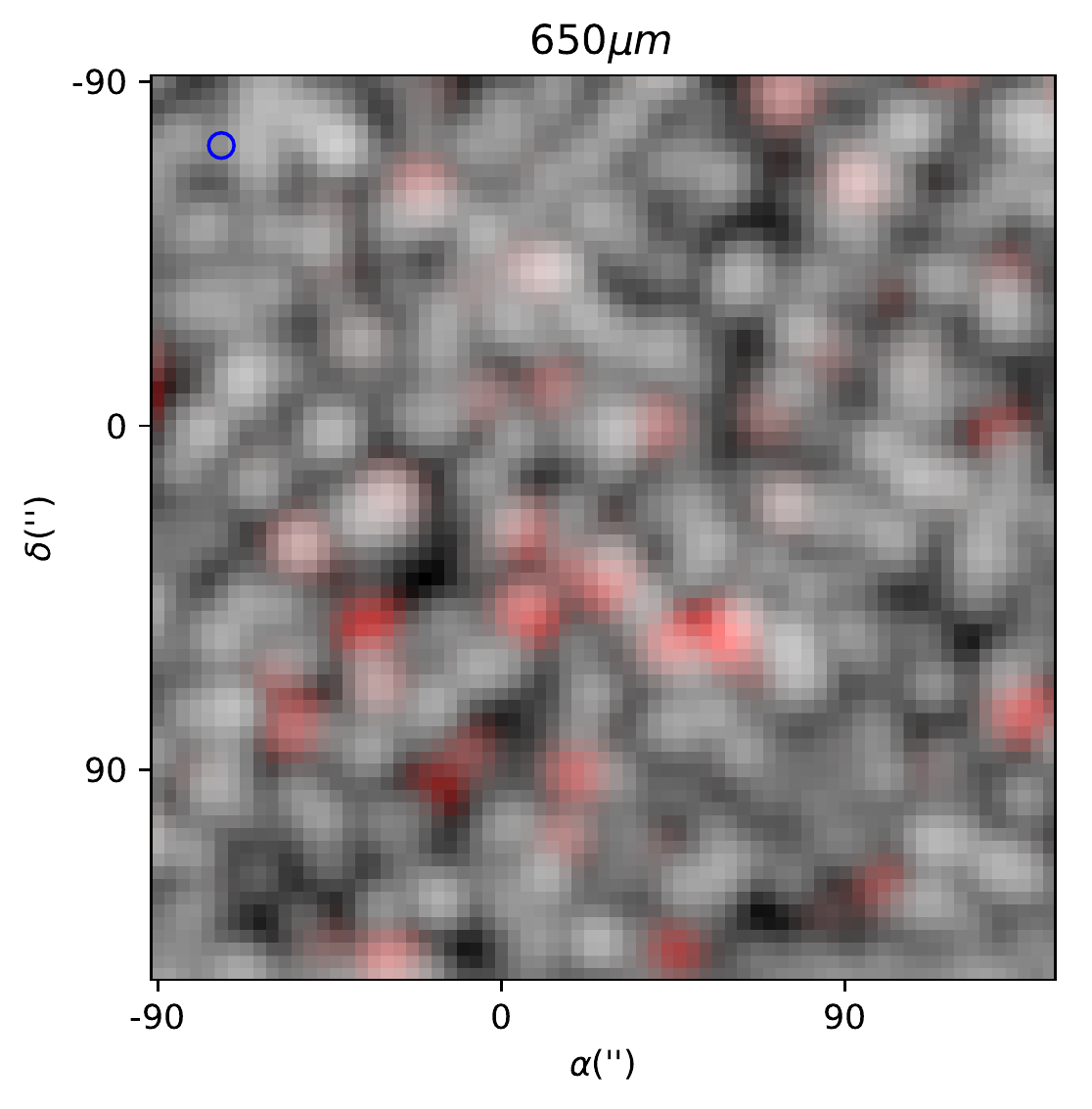}
\includegraphics[width=0.5\textwidth]{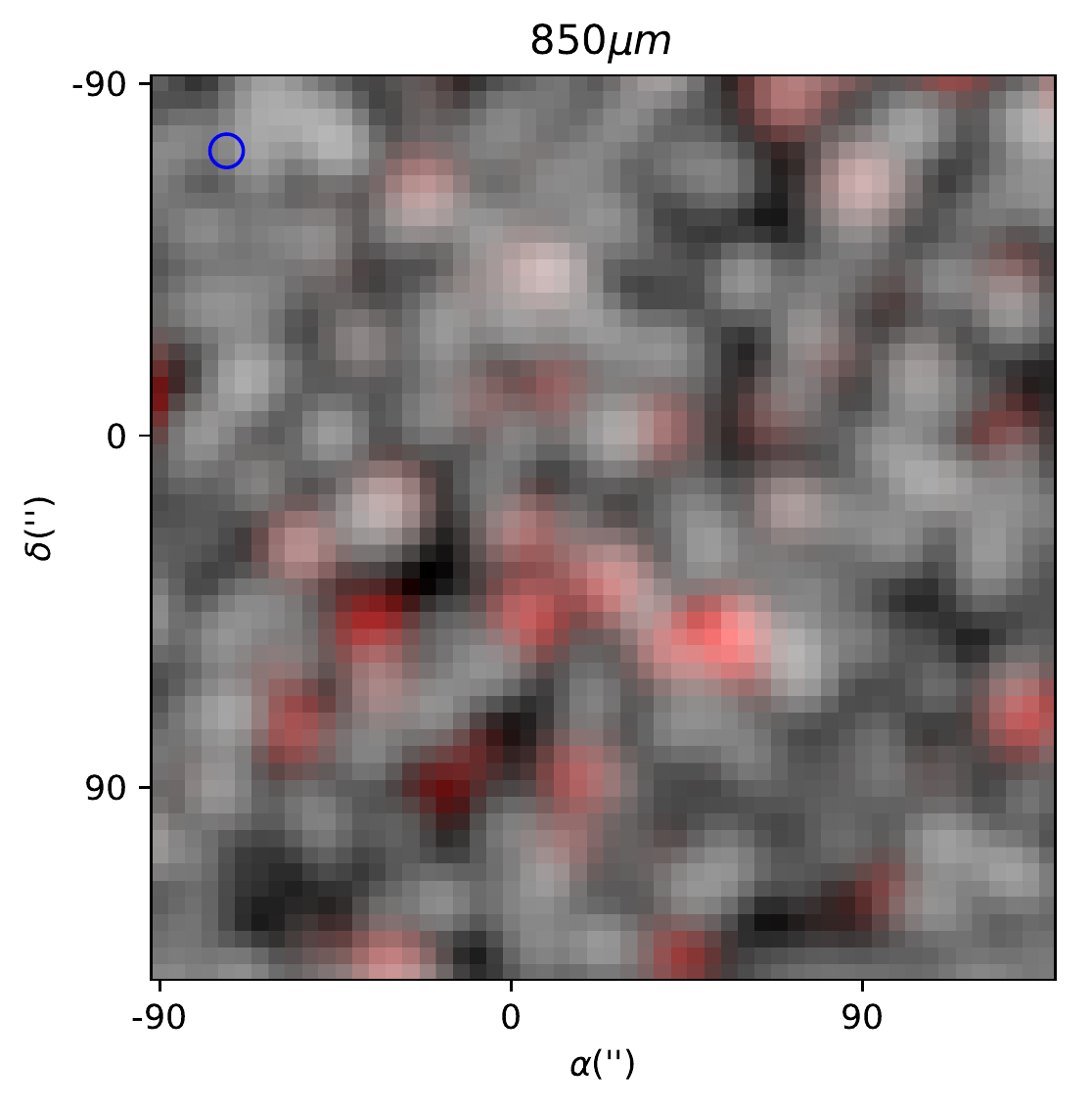}\\
\includegraphics[width=0.5\textwidth]{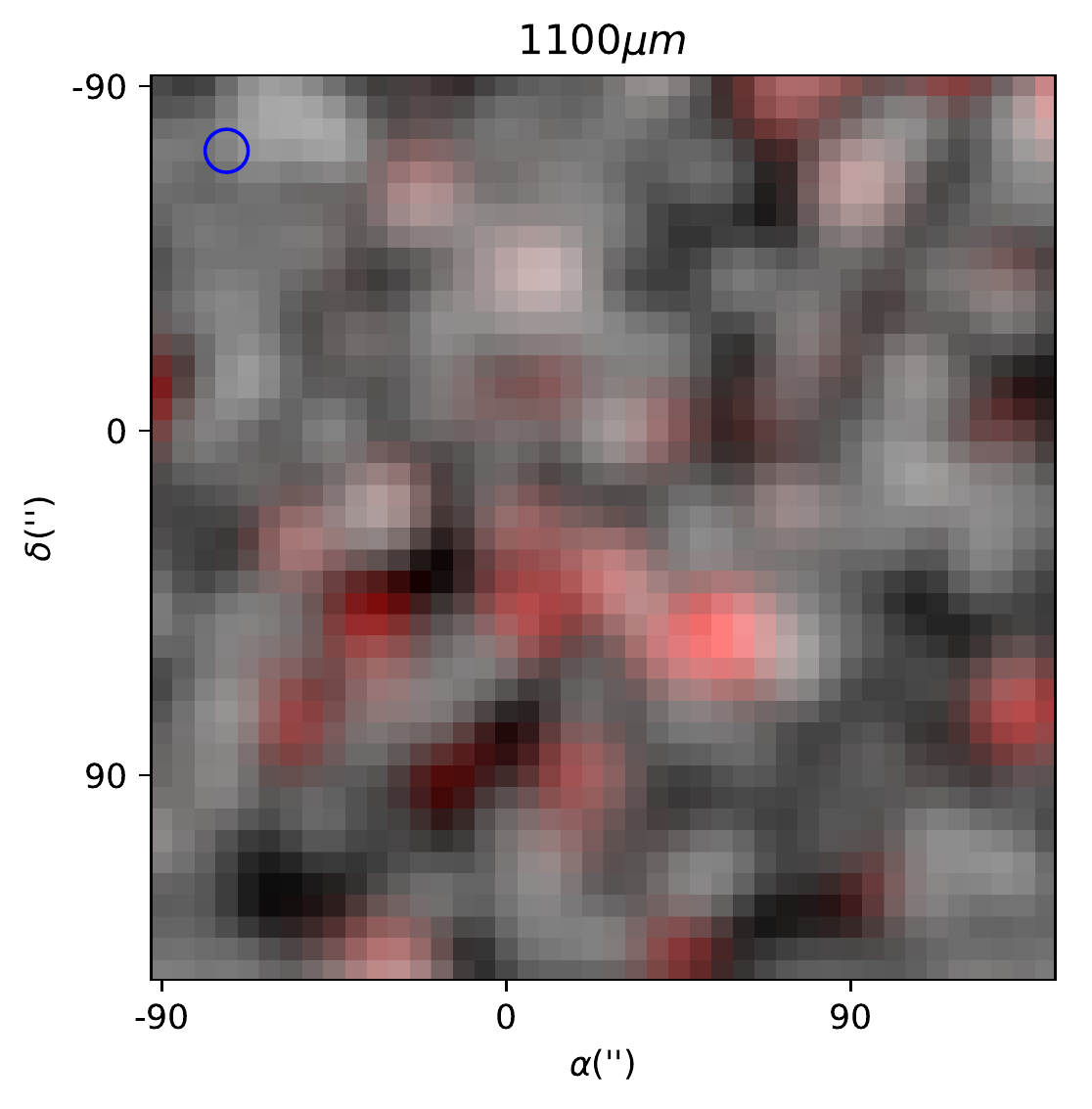}
\includegraphics[width=0.5\textwidth]{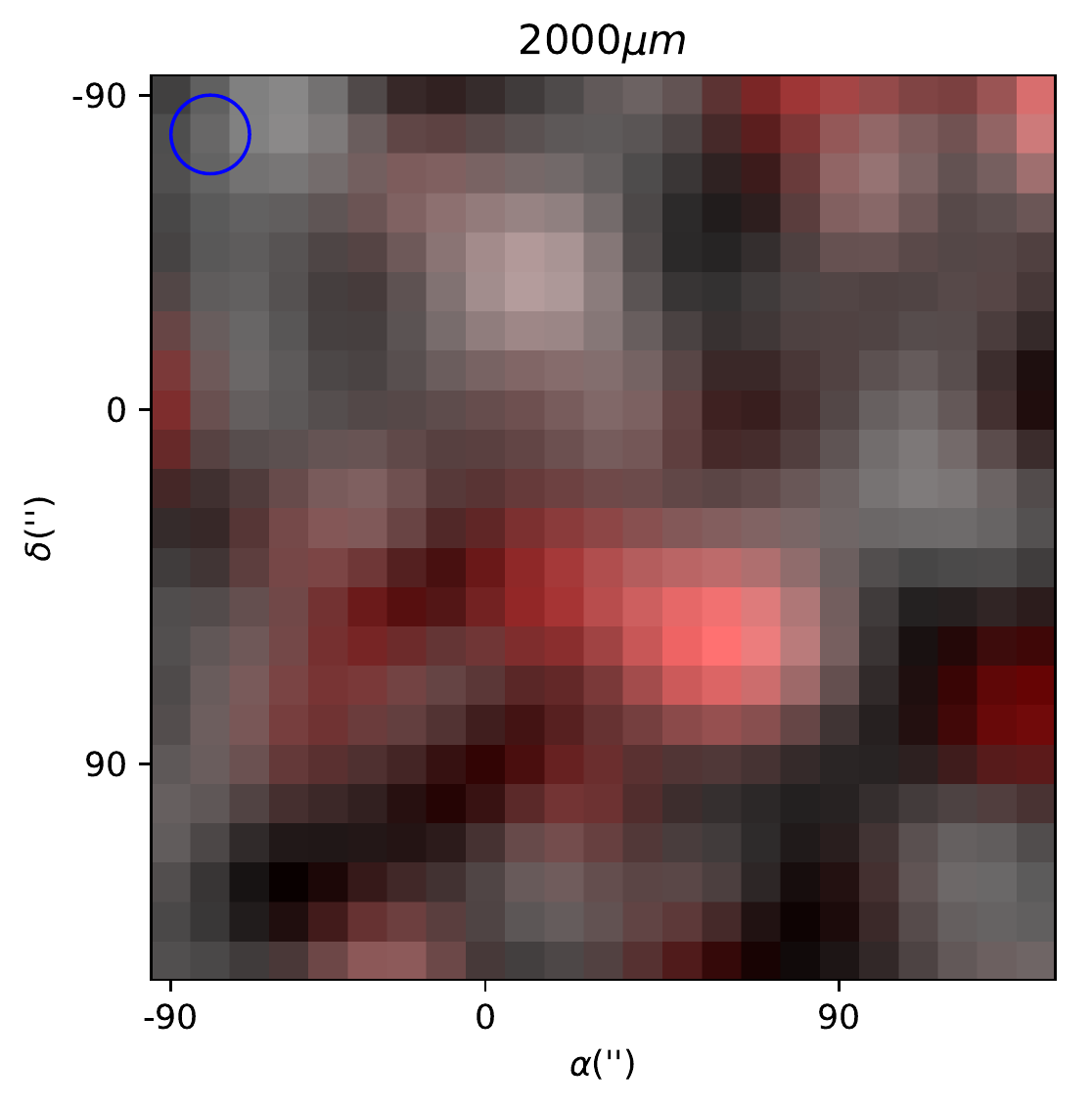}
\caption{Model maps for 650, 850, 1100 and 2000 $\mu m$.
Details are the same as on fig.~\ref{fig:maps1}.}
\label{fig:maps2}
\end{figure*}

In order to demonstrate the prospects of observation of lensed sources with the Millimetron mission we plot model maps of a small area for
eight wavelengths from 70 to 2000$\mu m$ (see figs~\ref{fig:maps1} and \ref{fig:maps2}).
The area of model maps is the same as the field of view of Millimetron.

In the process of creating the maps we accepted the following simplifications: sources were considered as point sources,
the beam of the instrument was Gaussian.
Intensity is in logarithmic scale.
Pixel angular size increases with wavelength.
So, for example, pixel size is 
0.6'' for 70  $\mu m$,
0.9'' for 110 $\mu m$,
2''   for 250 $\mu m$,
2.8'' for 350 $\mu m$,
4''   for 500 $\mu m$.

In order to visualize the contribution of lensed sources we used the following approach.
Let us consider a pixel that has a flux in RGB channels $R_{old}$, $G_{old}$ and $B_{old}$.
If we have to add to this pixel the flux $S_\nu$ of an unlensed source, we add it the following way:

\begin{eqnarray*} R_{new}=R_{old}+S_\nu,\\ G_{new}=G_{old}+S_\nu,\\ B_{new}=B_{old}+S_\nu
\end{eqnarray*}

The flux of a lensed source with magnification $\mu^{*}_{obj}$ is added the following way:

\begin{eqnarray*} R_{new}=R_{old}+S_\nu\times\\
(1+(\mu_{obj}-\mu_{min})/(\mu_{max}-\mu_{min})*2),\\
G_{new}=G_{old}+S_\nu\times\\
(1-(\mu_{obj}-\mu_{min})/(\mu_{max}-\mu_{min}))   ,\\
B_{new}=B_{old}+S_\nu\times\\
(1-(\mu_{obj}-\mu_{min})/(\mu_{max}-\mu_{min})),
\end{eqnarray*}
where $\mu_{obj}=min(\mu^{*}_{obj},\mu_{max})$, $\mu_{min}$ -- is the minimum value of the lensing coefficient,
$\mu_{max}$ -- the maximum value.

It is obvious that for an unlensed source the magnification coefficient is equal to 1.
Let us consider it as a minimal possible value $\mu_{min}$.
In case of a point source magnification can sometimes reach very high values.
But real sources have finite angular size that limits maximum lensing magnification and the probability of precise alignment of source and lens is quite low.
As will be shown in the following text the amount of sources with magnification greater than 2--3 is relatively low.
That is why in model maps $\mu_{max}$ was set to 3 in order to highlight most of the lensed sources, not only strongly lensed ones.

From the model maps on figs \ref{fig:maps1} and \ref{fig:maps2} it can be clearly seen that with increase of wavelength
the contribution of lensed sources increases while angular resolution decreases.
This effect can be explained by the growing contribution of distant sources.
It was already noted in numerous papers, see, e.g. \cite{2007MNRAS.377.1557N, 2010Sci...330..800N, 2011A_and_A...529A...4B} and references therein.
The reason behind it is the negative K-correction which means that for wide range of redshifts $z\approx1-4$ flux does not depend on redshift.

\subsection{Number counts of gravitationally lensed sources}

\begin{figure*}
\includegraphics[width=0.5\textwidth]{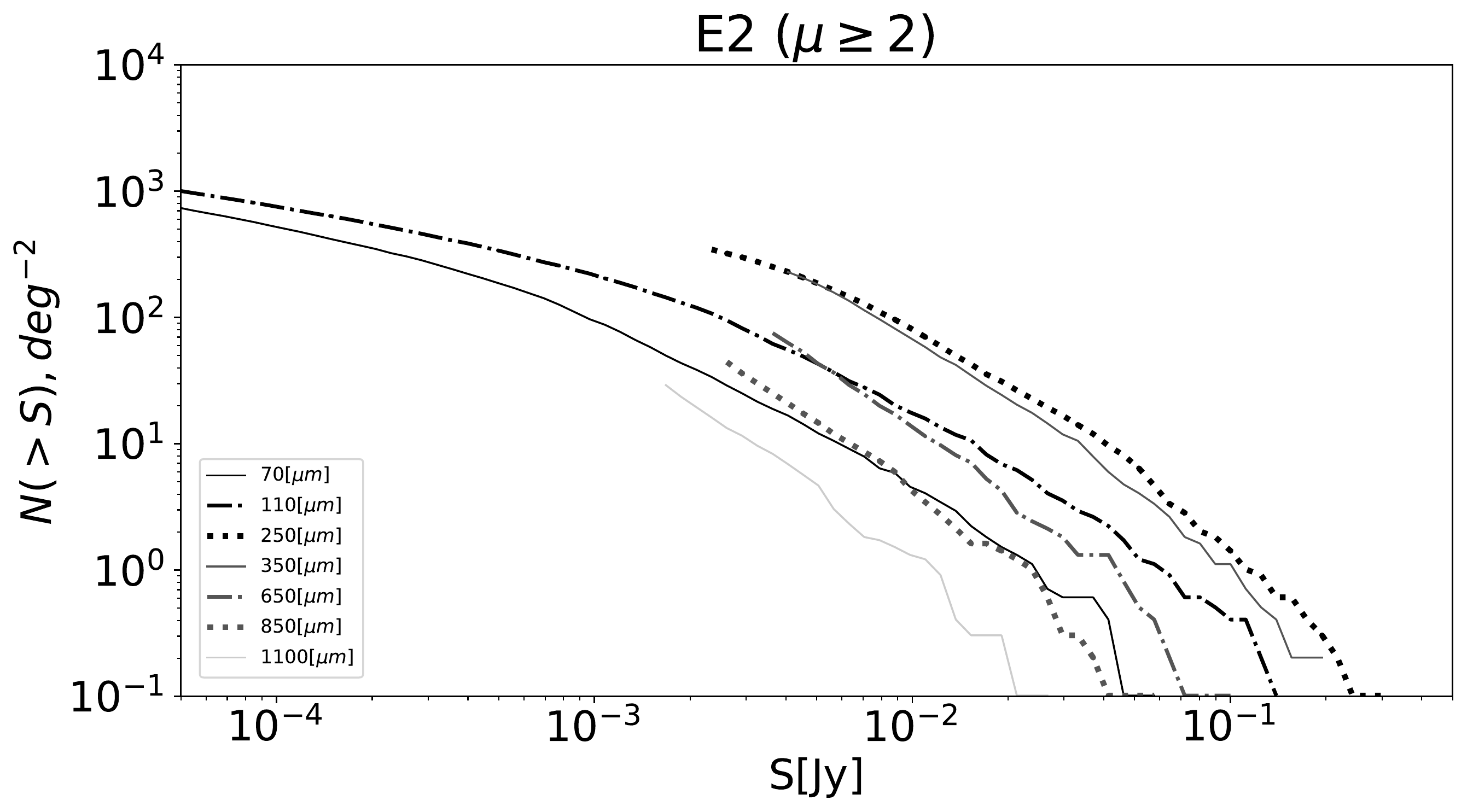}
\includegraphics[width=0.5\textwidth]{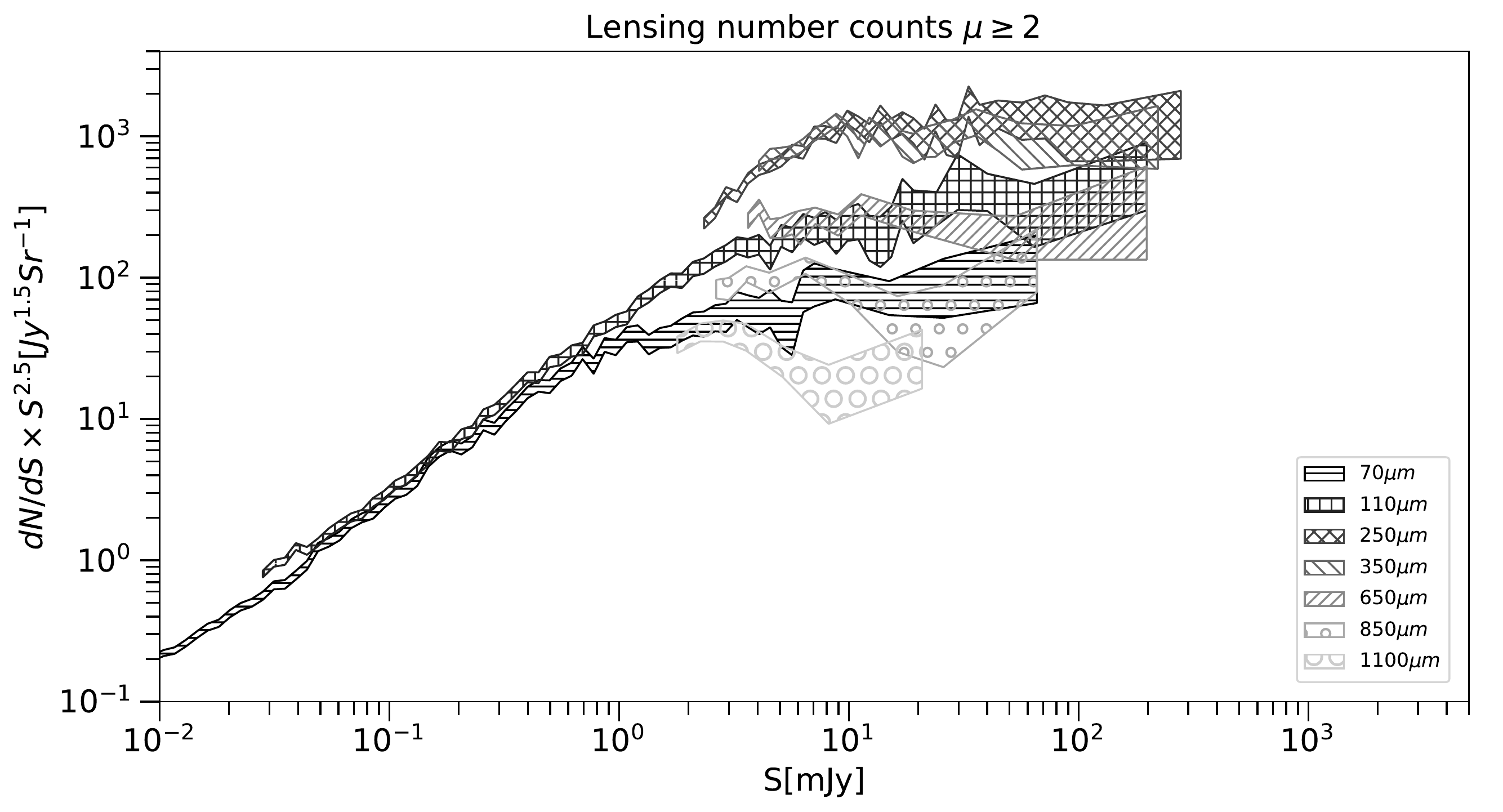}
\caption{Number counts of sources with magnification coefficient $\mu\geq2$ for the following wavelengths: 70, 110, 250, 350, 650, 850, 1100 $\mu m$.
Left~-- integral number counts $N(S)$; right - differential number counts.}
\label{fig:int_lensed_counts}
\end{figure*}

In order to approach the aforementioned scientific tasks and to create a draft of the observational program it is necessary to estimate the
amount of lensed systems available for observation on different wavelengths.
We built integral and differentional number counts of lensed sources with $\mu \geq2$ that are shown  on fig.~\ref{fig:int_lensed_counts}.
For example, the amount of sources with $\mu \geq2$ and flux greater than $\geq 1$mJy is $\sim 1000$ at millimeter and submillimeter wavelengths.

\begin{figure*}
\centering
\includegraphics[width=0.7\textwidth]{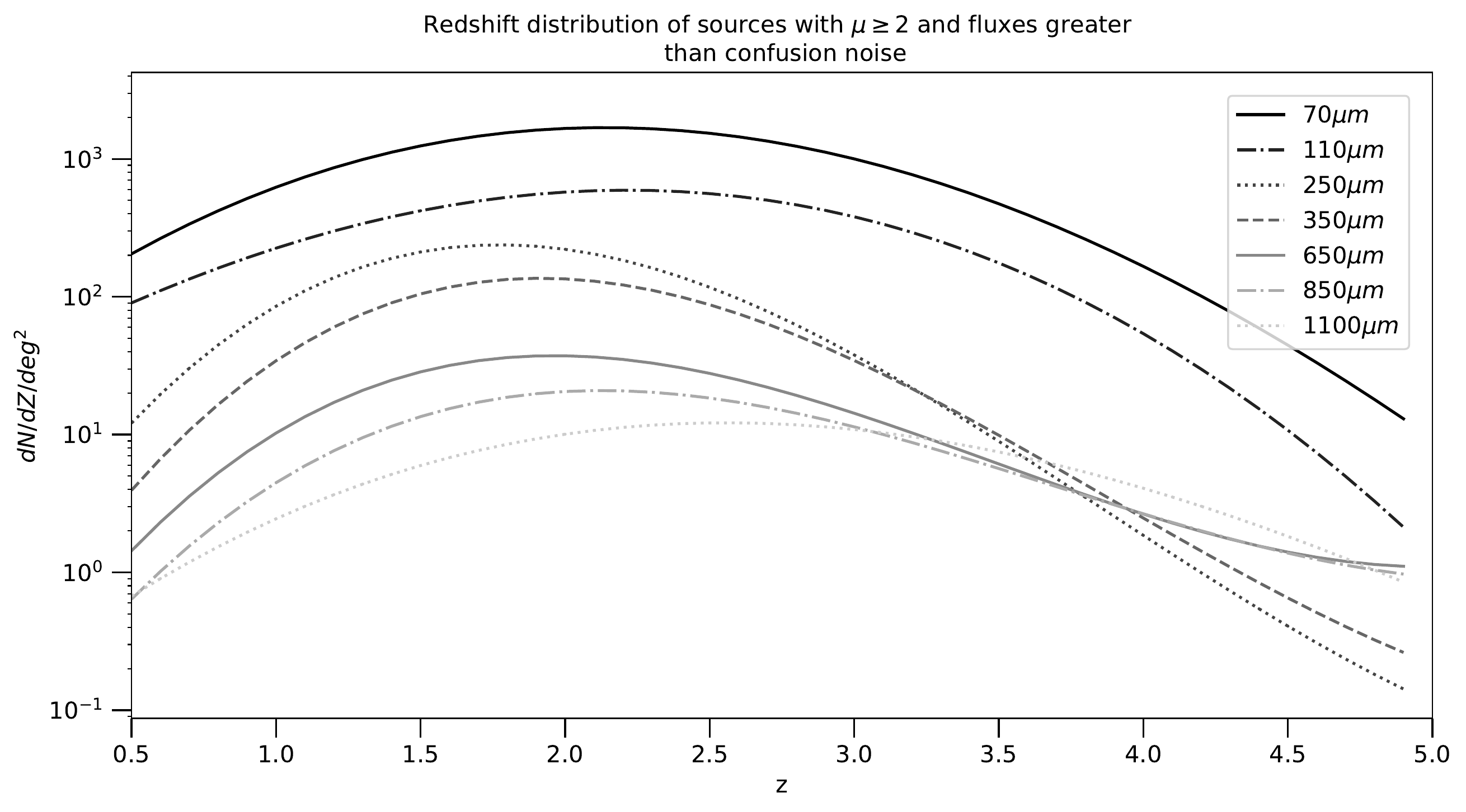}
\includegraphics[width=0.7\textwidth]{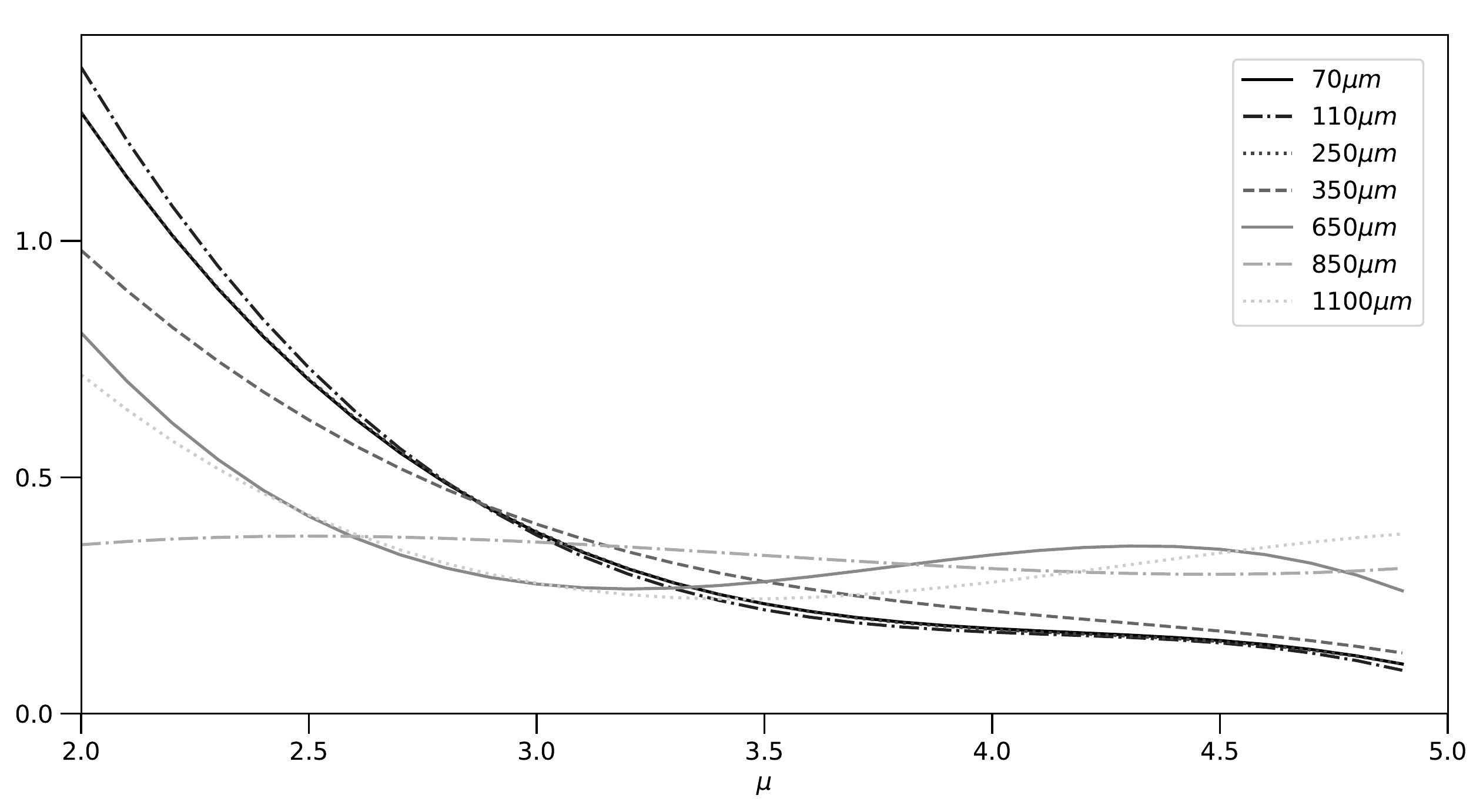}
\includegraphics[width=0.7\textwidth]{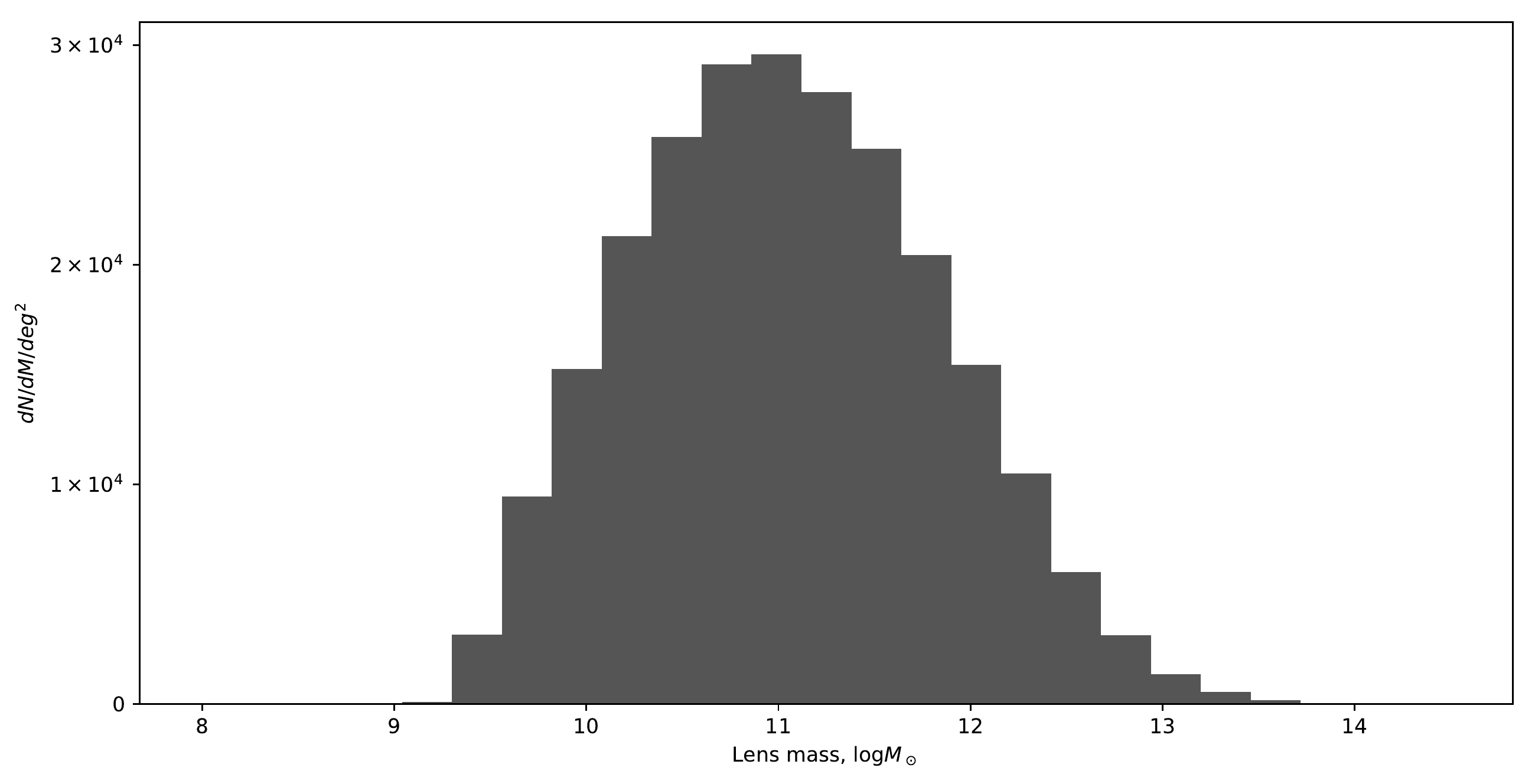}
\caption{Upper panel~-- redshift distribution of gravitationally lensed sources with magnification coefficient greater than $\mu\geq2$
for the following wavelengths: 70, 110, 250, 350, 650, 850, 1100 $\mu m$.
Central panel~-- magnification coefficient distribution of sources (normalized).
Lower panel~-- mass distribution of lenses with $\mu \geq 2$.}
\label{fig:redschift_distribution}
\end{figure*}

In order to estimate the amount of potentially observable lensed sources with $\mu\geq2$ on various redshifts, even at $z\geq2$, we need to know their exact 
redshift distribution.
Such distribution is shown on figure~\ref{fig:redschift_distribution} on the upper panel. 
The number of lensed sources grows up to $z\sim$1.5 on all wavelengths.
At higher redshifts on far infrared wavelengths it decreases rapidly with redshift, while on submillimeter wavelengths more
gradual decline can be seen due to the negative K-correction.

The normalized distribution of lensed sources with $\mu \geq 2$ by the lensing coefficient  is shown on the central panel of figure~\ref{fig:redschift_distribution}.
From this distribution it can be clearly seen that most of the lensed sources have magnification coefficients between 2 and 5.
This is in agreement with the results of Hershel observatory that bright lensed sources with $S_{500\mu m}>100$mJy have average magnification coefficient 
$\mu \approx 6^{+5}_{-3}$ \citep{2013ApJ...779...25B}.

In order to estimate the expected typical angular distance between lensed images of the object we need the
information about the mass distribution of the lenses.
This distribution is shown on the lower panel of figure~\ref{fig:redschift_distribution}. 
It can be seen that maximum of the distribution is about $10^{11}M_{\odot}$.
For a lens with such a mass angular distance between images will be about one angular second and that is comparable with the Millimetron angular resolution
(see table \ref{tab:millimetron_detectors}).
It is important to note that lensed DSFG galaxies that were detected on the SPT (South Pole Telescope) and were targets of
follow up high resolution ALMA observations had angular separation between lensed images about $\sim$1-2 angular seconds (see, e.g., \citealt{2013ApJ...767..132H}).
That value is in agreement with the maximum of the mass distribution of lenses.
The predicted mass distribution leads to the conclusion that half of lensed sources will have angular separation of images larger than 1 angular second.

\section{Scientific tasks and prospects of their solving}
\label{sec:science_task}
Let us now consider most actual scientific tasks that can benefit from observations in far IR, submillimeter and millimeter wavelengths.

\subsection{Independent determination of the Hubble constant}
\label{subsec:Hubble_constant}

As was already discussed above, the determination of the Hubble constant is one of the important scientific task because
this constant defines modern expansion rate of the Universe, its age, size and critical density.
Now there is a discrepancy at $4.4\sigma$  level between the value obtained by the Hubble Space Telescope with the ''distance ladder''
\citep{2019ApJ...876...85R}
($H_0=74.03\pm1.42$(km/s)/Mpc)  and the value obtained by the Planck mission in
$\Lambda CDM$ model by the measurements of the CMB \citep{2018arXiv180706209P} ($H_0=67.4\pm0.5$(km/s)/Mpc).
Data obtained from analysis of baryonic acoustic oscillations together with data on first type supernova (SNe Ia)
give a result close the value obtained by Planck observatory $H_0 = 67.3 \pm 1.0$ (km/s)/Mpc \citep{2017MNRAS.470.2617A}.

The $H_0$ value derived from megamaser observations
$H_0 = 66.0 \pm 6.0$ (km/s)/Mpc \citep{2016ApJ...817..128G}
and in the Carnegie-Chicago Hubble Program
$H_0 = 69.8 \pm 0.8 (\pm 1.1\% stat) \pm 1.7 (\pm 2.4\% sys)$ (km/s)/Mpc \citep{2019ApJ...882...34F}
are also close to this value.

So the results of recent observations show signs that for local Universe and the distant Universe the value of the Hubble constant might differ.
If that is really the case, then the $\Lambda CDM$ model should be modified.
E.g. one should consider dynamical dark matter, non-zero curvature, larger number of effective relativistic particles etc.
In order to figure out the nature of the discrepancy of $H_0$ independent methods of its measurement should be used.
One of such methods is the measurement of the so-called delay distance between multiple images of a bright variable lensed source, like quasar or a supernova.
This approach was developed by \cite{1964MNRAS.128..307R}.
The delay distance is reversely proportional to the value of the Hubble constant.
Such an approach can be used not only to constraint the value of this constant but also another cosmological parameters.
The delay distance is determined by the measurement of delay of photons from different images of the lensed source.
But the delay itself is not sufficient to determine the $H_0$ with good enough accuracy (see, e.g., \citealt{2011AstL...37..233L}).
\cite{2014ApJ...788L..35S} showed that for one lensed object with measured time delay and some additional observational data the Hubble constant can be derived with
7--8\% accuracy.
Today the Hubble constant is measured with 2.4\% accuracy by the analysis of six lensed quasars and is found to be 
$73.3^{+1.7}_{-1.8}$ (km/s)/Mpc \citep{2019arXiv190704869W}.
So, taking this result into account, the tension between the estimations, the discrepancy between estimations of the $H_{0}$ by the analysis of observational data of young and local Universes is 5.3 $\sigma$.
But in order to solve the problem of the discrepancy between the estimations of the Hubble constant from various experiments and to
obtain the information about the dark energy the 1\% accuracy is required.
For such accuracy one needs to have detailed information and the time delay between images for about 40 gravitationally lensed systems
(see, e.g. \citealt{2019arXiv190704869W} and references therein).

\begin{figure*}
\centering
\includegraphics[width=0.7\textwidth]{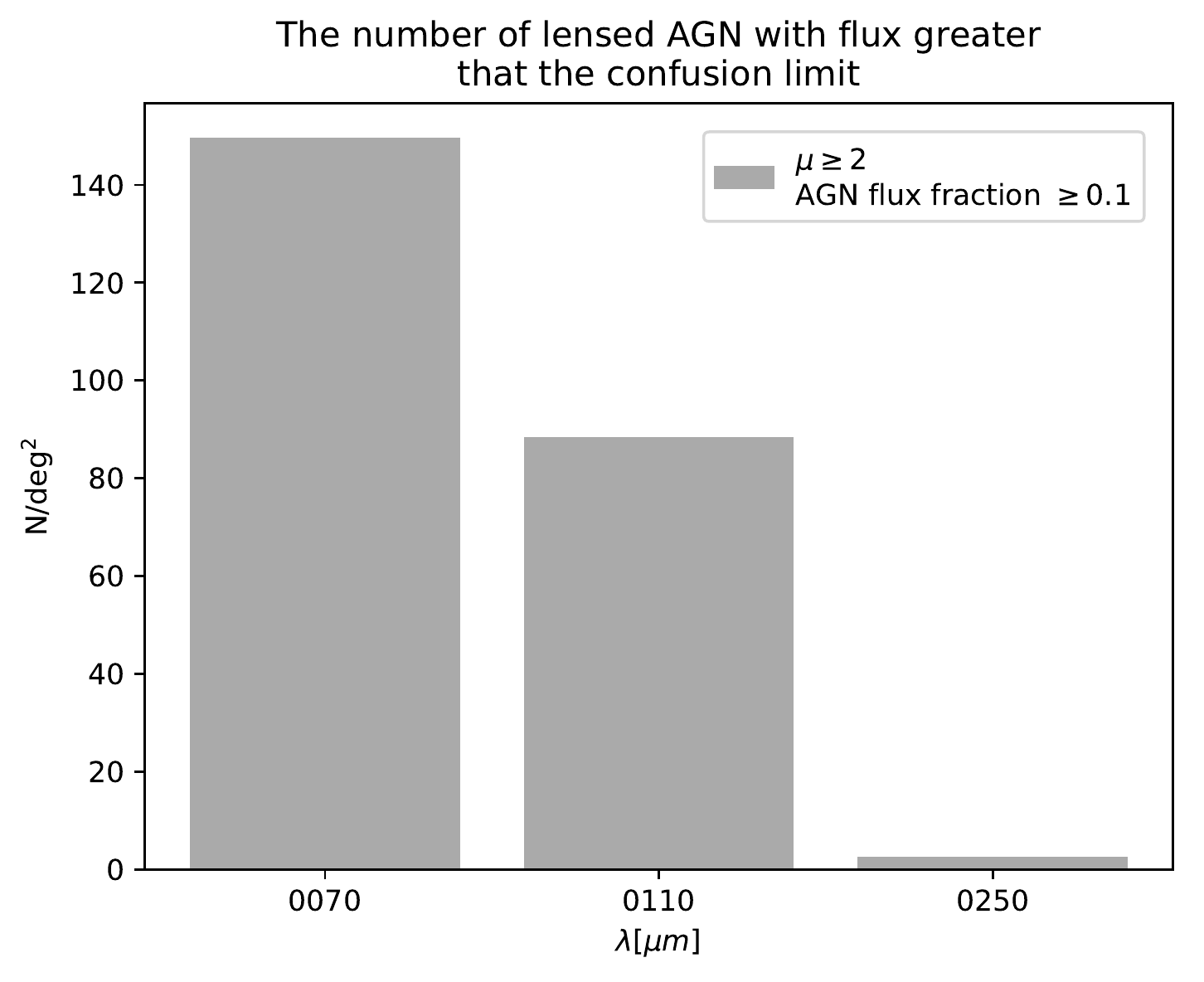}
\caption{The histogram showing the number of lensed AGN on 70--250$\mu m$ with flux greater than the confusion limit and the lensing coefficient $\mu \geq 2$.}
\label{fig:agn_counts}
\end{figure*}

All the aforementioned estimations of the Hubble constant were obtained by means of observation in the visible, near infrared or radio wavelength ranges.
The Millimetron space telescope will be able to provide additional information about gravitationally lensed systems in far infrared and 
submillimeter wavelength ranges.
According to the table~\ref{tab:millimetron_detectors} the resolution of the telescope at the shortest wavelengths will be about 1--2 angular seconds
and the sensitivity of the wideband photometry will reach 0.01 mJy. 

In order to estimate the number of potentially observable gravitationally lensed systems that can be used for the estimation of $H_0$
we need to know the amount of AGN with magnification coefficient $\mu \geq 2$  and the distribution of masses of lenses to estimate the
angular distance between images.
The number of quasars with $\mu \geq 2$ at 70--250$\mu m$ with flux greater than the confusion limit is shown on fig.~\ref{fig:agn_counts}.
This histogram is created for the objects with the AGN flux fraction greater than 0.1.
It can be seen that about 140 AGN with $\mu \geq 2$ can be detected on the sky area of one square degree in the wavelength range 70--250$\mu m$.
The histogram of the distribution of the masses of lenses in shown on figure~\ref{fig:redschift_distribution}.
From this distribution a conclusion can be made that about a half of lensed sources will have angular separation of images greater than one angular second
and about $\sim15\%$ greater than 2 angular seconds.

According to these estimates in order to register 40 lensed AGNs with angular separation between images greater than 2 angular seconds the survey
of 2 square degree area is required. The scan speed of the Millimetron space telescope will be about 0.005 sq. deg. per minute, so the observation
time required is about 7 hours.

Another promising approach for the task considered is to observe previously detected AGNs in different wavelengths that have 
measurements of time delay between images of a variable source.
%The most promising targets to search for gravitationally lensed systems are the massive galaxy clusters that have estimations of the gravitational potential,
%accurate mass and distance measurements, in another words clusters that have accurate model of the lens.
Observations of AGN in far infrared wavelength range will provide additional information that together with data at other wavelength ranges will
make the independent estimation of the Hubble constant with required accuracy possible in order to determine the difference between this value in
the local and distant Universes.
It must be stressed that for this task, like for all other scientific tasks dedicated to the study of gravitationally lensed sources,
besides the photometric observations with high sensitivity and fine angular resolution spectroscopic data with high spectral resolution are required
in order to derive spectroscopic redshifts of lenses and lensed sources.

\subsection{Distant galaxies with high amount of dust and active star formation}
\label{subsec:DSFG}

Dusty Star Forming Galaxies are most massive and have extreme star formation rates,
their typical stellar mass and SFR are
$\sim 10^{10} M_{\odot}$ and 
$\sim 100 M_{\odot}$yr$^{-1}$
, respectively
(see, e.g., review by \citealt{2014PhR...541...45C} and references therein).
Due to such extreme star formation these galaxies host large amounts of dust. More than 95\% of the radiation comes from hot young stars surrounded by
dust that re-emits the UV radiation in far IR (see, e.g. \citealt{1996ARA&A..34..749S,2002PhR...369..111B}).

In the local Universe such galaxies are quite rare, while at $z\sim 2$ they are about 1000 times more frequent and become the 
main contributors to the total star formation in the Universe \citep[see, e.g.][and references therein]{2014PhR...541...45C}.
Probably all the DSFG are the result of merging of galaxies or are massive gas rich disc galaxies in stage
of intense star formation (see, e.g. \citealt{2012MNRAS.425.1320I,2016ApJ...833..103H}).

On high redshifts ($z>2.5$) the volume density of DSFG is unknown. The same can be said about their contribution to the global star formation rate
in the Universe. The knowledge about the properties of DSFG in early Universe is crucial for understanding of the mechanisms of galaxy formation
and star formation processes in early Universe and formation of dust (supernova, AGB stars or growth of dust particles in interstellar medium).

Today there are 77 discovered candidates in bright DSFG in HerMES Large Mode Survey and Hershel Stripe 82 Survey that covers 372 sq. degrees
\citep{2016ApJ...823...17N}.
List of about a hundred of potential candidates can be found in \cite{2013Natur.495..344V}.

As was mentioned above, the best strategy to search for submillimeter lensed galaxies on high redshifts is to perform wide area shallow surveys 
due to the negative K-correction.

The expected amount of lensed galaxies with $\mu\geq2$ on a 1 square degree area is expected
to be $10^2$--$10^3$ for different wavelengths (see number counts curves on fig.~\ref{fig:int_lensed_counts}).
Because of the high sensitivity the scan speed will reach 0.05 square degrees per minute at $\lambda>110\mu m$ and an order of magnitude lower at 
shorter wavelengths.
The area of one square degree can be covered by Millimetron in $\sim20$ or $\sim200$ minutes depending on the exact wavelength.
According to the aforementioned estimations 90\% of the strongly lensed objects will be distant DSFG.
Final determination of the object's type will require the SED analysis. The most promising strategy to enlarge the sample of DSFGs with $z\sim1-2$ and to
search for weak distant sources is the observation of galaxy clusters in far infrared and submillimeter wavelengths (see, e.g. \citealt{2013ApJ...769L..31Z}).
Galaxy clusters are mainly populated by early type galaxies that do not have significant emission in submillimeter wavelengths, so clusters are in fact 
act as transparent lenses.
Galaxy clusters from ``The Heshel Lensing Survey'' and ``HST Frontier Field Coverage'' will be included in the observation program.
These fields will be scanned two orders of magnitude faster compared to the Hershel obsevatory.
Moreover, many of these clusters have the known distribution of gravitational potential, accurate mass and distance measurements, in another words one have accurate model of the lens.

\subsection{First galaxies}
\label{subsec:LAE}

Another important task for cosmology is the study of the earliest galaxies that probably played the key role in the reionization of the Universe and
in the radiation heating at high redshifts  $z \approx 7 - 12$ (see, e.g. \citealt{2016MNRAS.463.4019K}).
Such heating prevents the accretion of baryons onto low mass haloes that causes the decrease of amount of low mass haloes containing stars.
The heating of interstellar medium suppresses the star formation in massive haloes, in another words leading to the decrease of star formation rate.
It is expected that masses, luminosities and metallicities in early galaxies were close to the values for the satellites of our Galaxy and Andromeda galaxy.
The researches of the dwarf galaxies of Local Group showed that formation of metals is possible after the 
formation of the galaxy or in the later period~\citep{2014ApJ...789..148W}. 
In order to find out which scenario took place in the first galaxies it is necessary to observe high redshift galaxies.

It is possible that the so-called Lyman-Alpha Emitters (LAE) are connected to the first galaxies.
These are extremely bright objects that emit significant portion of their energy in the $L\alpha$
line due to the scattering of the UV radiation of the central source (or the star formation region) in the surrounding envelope 
consisting of neutral or mildly ionized hydrogen.
These objects are distributed in a wide range of redshifts up to $z\sim11$, which points to the fact that they play a crucial role in the reionization of the 
Universe.
Observations of several hundreds of LAE candidates lead to the conclusion that they are low mass galaxies with active star formation and that they
can be considered as progenitors of dwarf galaxies in local group (see, e.g. \citealt{2018ApJ...864..145H,2018PASJ...70S..15S}).

LAE on high redshifts will be observed in the Paschen and Brackett series that fall in the wavelength range of the Millimetron telescope.
Despite the fact that their intensity is low, the sensitivity of the Millimetron detections ($\sim10^{-22}Wm^2$) will make their detection possible.
The magnification of these objects due to gravitational lensing makes their detection more likely.

The preferred targets for registration of LAE's emission are massive galaxy clusters that are listed in \cite{2019arXiv190302002C}.
To solve one of the key task the Millimetron scientific program will include observation of about a hundred of such clusters. 
Considering the results obtained by \cite{2019arXiv190302002C} we expect to detect several hundred of LAE candidates.
The selected candidates will become targets of spectroscopic observations in submillimeter and millimeter wavelength ranges.
The lines of interes are [CII] 158 $\mu m$, [OIII] 88 $\mu m$, [OI] 63 $\mu m$ and CO lines.

In the recent years there was a significant progress in the spectroscopy of the lensed galaxies, including galaxies lensed on galaxy clusters
\citep[see, for example,][]{2014MNRAS.445.3200S,2015MNRAS.450.1846S}.
Recently \citealt{2013ApJ...771L..20K,2019MNRAS.487L..81L,2019ApJ...881..124M}  showed that lensed LAE do not show [CII] emission that is clearly observed in the submillimeter galaxies.

The Millimetron observatory will increase the LAE sample and analyze the emission in this line with high sensitivity and resolution.
That will allow to validate the hypothesis of existence of several kinds of LAE galaxies.

\subsection{Another scientific tasks}

It is a well known fact that numerical simulations following the standard $\Lambda CDM$ model reproduce the large-scale structure fairly well, but
the amount of observed galaxies is significantly lower that modeled.
This is the definition of the so-called satellite problem.
That is why the knowledge about the exact number of low mass galaxies is crucial for the choice of the cosmological model and the dark matter model.
The strong gravitational lensing gives the most straight approach of determination of the dark matter distribution on sub-galactic scale.
\cite{1998MNRAS.295..587M} proposed to use the observed ratio of fluxes of lensed images of quasars to limit the number of massive structures (subhalo) in galaxies that
work as lenses.
\cite{2002ApJ...572...25D} used relatively small sample of seven lensed quasars to derive statistical limitations on the amount of the projected mass contained
in substructures.
They showed that in general it corresponds to the results of numerical simulations in the $\Lambda$CDM model.
The same problem was also considered in \cite{2002ApJ...567L...5M}.
At the same time another analysis by \cite{2009MNRAS.398.1235X,2015MNRAS.447.3189X} showed that the population of substructures predicted  by numerical methods is 
insufficient to explain the observed anomalies in the ratio of fluxes of lensed images. 
These anomalies can be probably explained by the presence of complex baryonic structures in lensing galaxies and also by a presence of 
dark subhaloes on the line of sight.
 
Aside from the analysis of anomalous flux ratios of lensed images there are other methods to detect substructures in strongly lensed sources.
One of such approaches is the ''gravitational imaging'' technique that uses the extended arcs and Einstein rings to detect and analyze the masses of 
separate subhaloes \citep[see, for example,][]{2005MNRAS.363.1136K,2009MNRAS.392..945V}.
This approach considers the substructures not as analytical clumps of mass but instead as linear pixel corrections of global lensing potential.
That is why it does not require any assumptions on the number of substructures and their density profile  and redshift.
To say more, this approach can easily discriminate between the presence of a substructure and smooth but complex mass distribution.
That is why this method is less prone to false detections.
But in contrast to the analysis of anomalous flux ratios in order to detect a low-mass subhalo observational data with high resolution and high flux 
dynamic range are required.

\section{Discussion}
\label{sec:discussion}
In this paper we have discussed the prospects of observation of gravitationally lensed extragalactic sources in far infrared and submillimeter wavelength ranges
by the next scheduled for launch space observatory with actively cooled main mirror that will perform observations in this wavelength range.
All the calculations were performed accordingly to the declared specifications of the observatory, in particular the cooled (T<10K) main 10m mirror
with high sensitivity detectors that will work in wavelength range 70 -- 2000 $\mu m$.
The number counts of lensed sources for different wavelengths were calculated.
We built the distributions of the lensed sources by redshift and magnification coefficient and the mass distribution of lenses.
We created model sky maps that illustrate the contribution of lensed sources. 
The prospects to solve actual cosmological and astrophysical tasks connected to the observation of gravitationally lensed systems were discussed.
Among them are the research of distant galaxies with active star formation and significant amount of dust (DSFG), 
first galaxies such as Lyman-Alpha emitters, independent determination of Hubble constant and other cosmological parameters (e.g. dark energy), subhalo
detection, dark matter distribution in galaxies and massive galaxy clusters.

\vspace{15mm}
This study was supported by the program KP19-270 ''Questions of origin and evolution of the Universe with ground based and space researches.''

\clearpage
\newpage
 
\bibliographystyle{rusnat}
\bibliography{bibliography}

\end{document}